# Remarks about the Zeeman splitting in quantum dots and single atom transistors

V.V. Fisun

*B.Verkin Institute for Low Temperature Physics and Engineering, National Academy of Sciences of Ukraine, 47 Lenin Ave., Kharkiv, 61103, Ukraine*

An empirical formula is proposed to calculate the Kondo temperature in quantum point contacts from the Zeeman splitting. A similar formula was used in the point-contact spectroscopy (PCS) method to obtain $T_K$ from the Zeeman splitting of the Kondo peak in the $dV/dI(V)$ characteristics of the classical Kondo alloys AuFe and CuMn [1]. Using this formula, it is possible to find the Kondo temperature taking into account the temperature of the experiment and Lande factor of the compounds.

The classical Kondo effect [2] manifests itself as a minimum in the dependence of electric resistance upon the temperature of dilute alloys of noble metals (Cu, Ag, Au, etc.) with paramagnetic impurities (d-metals: Ti, V, Cr, Mn, Fe, Co, Ni ) Below the temperature of the minimum $T_{\min}$, the electric resistance grows by the logarithmic law $\rho \propto a - b\ln T$ as the temperature decreases. This phenomenon is of quantum nature and is determined by the antiferromagnetic exchange interaction between the conduction electrons of a nonmagnetic metal and the magnetic impurities – atoms with open *d*- or *f*-shells.

In an external magnetic field *H*, the dependence *R(T)* has a maximum with the characteristic $T_{\max}$ in the interval $0 < T_{\max} < T_{\min}$, which appears against the general background of negative magnetoresistance. The maximum is due to the Zeeman energy $E = -g\mu_B H$. It is observed with $T_K$ of several $K$.

The recently developed technologies of making nanometer-scale structures have possibility to studying the properties of single-electron (QD) [3] and single-atom (SAT) [4] transistors, quantum point-contacts (QPC) [5] at ultralow temperatures. The main effect in a single-electron transistor is the quantization of charge on a very small island. This leads to Coulomb gaps in the energy spectrum of this island and to a blockade of transport. Quantum dots are small electronic devices, which confine a well-defined number of electrons, N. The total spin is zero or an integer for even N, and half-integer for odd N. The latter case is the canonical example of the Kondo effect when all electrons can be ignored, except for the one with the highest energy; that is, the case of a single, isolated spin, S = ½. At low temperatures the correlations of the electrons on the QD and within the banks to which the QD is connected by tunnel junctions become important.

Consequently, the Kondo effect in QPC and QD on GaAs/AlGaAs heterostructures containing two-dimensional electron gas (2DEG) differs from the classical case in that it has no magnetic impurities.

Classical Kondo alloys with $T_K$ of several Kelvins were investigated using tunnel [6] and PC [7] spectroscopy. The differential resistance in the point contacts of the Kondo alloys has a peak at $V = 0$ (the so called «zero bias anomaly» - ZBA) In the ZBA region the differential resistance increases with the decrease of *V* by the logarithmic law $R_{diff} \approx a - b\ln V$. It looks as if the logarithmic dependence of the resistance growth in $R_{bulk}(T)$ has moved to $R_{diff}(V)$ in point contacts. In the magnetic field the Kondo peak is split by $V^{P-P}$ (Zeeman splitting). The splitting produces two peaks located symmetrically with respect to $V = 0$. There are just two peaks (splitting) because the spectra of the dependences *dV/dI(V)* and

$dI/dV(V)$ are presented in bipolar coordinates as a function of the applied voltage. The energy $V^{P-P}$ is proportional to the Zeeman energy determined by the external magnetic field.

A similar ZBA was observed in QPC and QD at a certain gate voltage. In that case the Kondo peak occurred in the conductance $dI/dV(V)$ rather than in the differential resistance $dV/dI(V)$, as it is observed in PCS. Like in PCS, the Kondo peak of QD is split in the external magnetic field [3,4,8-12]. In the papers cited the g-factor was calculated by the formula $eV^{P-P} = 2g\mu_B H$ using $V^{P-P}$ measured for the Zeeman splitting of the Kondo peak. The calculated g-factor exceeds the corresponding value for the bulk material (g=-0.44 for GaAs). A similar discrepancy was earlier observed in PCS studies on classical Kondo alloys at g=2 [13] for classical Kondo alloys.

The splitting of the Kondo peak in classical Kondo alloys was analyzed in a number of theoretical studies [14, 15]. In [15] $V^{P-P}$ of theoretical spectra exceeds the Zeeman splitting $2g\mu_B H$ at g=2, but it is still lower than experimental values.

We believe that, like in PCS [13], the higher g-factor obtained from $V^{P-P}$ is connected with the calculation by the formula $eV^{P-P} = 2g\mu_B H$ which does not allow for the experimental and Kondo temperatures.

We solved this problem while investigating the Kondo size effect in the break-junction type of point contacts [16] on dilute alloys CuCr, CuMn and AuFe in an external magnetic field [1]. When the Kondo peak is split by the magnetic field $V^{P-P}$ is dependent on both the Zeeman energy and the temperatures $T$ and $T_K$. For the CuMn and AuFe alloys these temperatures are taken into account in our empirical formula (see Eq. (1) in [1]). The formula agrees well with the Zeeman splitting in [13]. In a certain magnetic field, a decrease in the point contact diameter causes $V_{\exp}^{P-P}$ to increase, which is connected with the growth of the Kondo temperature due to the size effect.

The physical sense of the empirical formula is evident from a comparison of the dependences $R(T)$ and $dV/dI(V)$ or $dI/dV(V)$ in an external magnetic field $H$. In the PCS and QPC cases the dependences $dV/dI(V)$ or $dI/dV(V)$ have similar $T_{\max}(H)$ in the form of two peaks at $\pm V_{\max}(H)$. The energy of the peaks is found as $V^{P-P} = 2V_{\max}$. The question arises as to what we can see in the dependences $dV/dI(V)$ or $dI/dV(V)$ when the temperature of the experiment is above $T_{\max}(H)$? It is obvious that no Zeeman splitting the Kondo peak will occur. Thus, the splitting in the dependences $dV/dI(V)$ or $dI/dV(V)$ is only possible when the temperatures of the experiment are in the interval $0 \leq T_{\exp} \leq T_{\max}(H)$. Now we can trace the behavior of $V_{\exp}^{P-P}$ when the temperature varies within this interval. Since, $eV \approx k_B T$, it is likely that $V_{\exp}^{P-P}$ will be the highest at $T = 0$ and will decrease as the temperature of the experiment approaches to $T_{\max}$. However, the PCS experiment on the CuMn and AuFe alloys [1, 13] show that in this interval $V_{\exp}^{P-P}$ increases with the temperature of the experiment. and Kondo, following our empirical formula (1) [1].

In case CuCr alloy, the dependence of $V_{\exp}^{P-P}$ with the temperature of the experiment is completely different from that in CuMn and AuFe. At T= 0.5 K $V_{\exp}^{P-P}$ is lower then $2g\mu_B H$ at $g = 2$, which follows reasonably from the above consideration because the temperature of the experiment and Kondo are nonzero. As the temperature of the experiment rises, $V_{\exp}^{P-P}$ decreases in comparison with $2g\mu_B H$ at $g = 2$. The dependence of the Kondo temperature of this alloy was found by Eq. (2) of Ref. [1]. We believe that the behavior of $V$ is similar in SAT too [17].

Let us again consider QPC for which $V_{\exp}^{P-P}$ exceeds $2g\mu_B H$ at the g-factor of a bulk. Like in PCS [1], we propose to use the empirical formula to describe the Zeeman splitting in QD, QPC and SAT. In this case the formula should be modified to take into account the temperature broadening $eV \approx k_B T$ for QD and QPC.

Like in Ref. [1], the formula proposed becomes:

$$eV^{P-P} = 2[g\mu_B H + k_B(T + T_K)] \quad (1)$$

where $V^{P-P}$ is the experimental value for splitting Kondo peaks (mV), g is the modulus Lande factor in bulk, $\mu_B$ =0.05789 mV/T is Bohr magneton, $H$ is the magnetic field (T), $k_B = 0.086$ mV/K is the Boltzman constant, $T$, $T_K$ are the temperature of the experiment and Kondo, respectively.

The formula (1) can be used only in a nonzero magnetic field of a sufficient value $g\mu_B H \geq k_B(T + T_K)$ when the Zeeman splitting are observed in the experiment.

Below we illustrate how the Kondo temperature can be found using the proposed formula. The results on the splitting of the Kondo peak are taken from Refs. [3,4]. This works illustrate the dependences $dI/dV$ (in units 2e²/h) of the Zeeman splitting of the Kondo peak in QPC and QD on the surface of the semi-conductive GaAs/AlGaAs and calculate the experimental g-factor. In [3] the g-factor in the fields $B \leq 3$ T is about $\approx 1.5$ times higher then the value for a bulk. In Fig. 2 (d) [3] the Zeeman splitting of the Kondo peak of GPC is shown in the region 0.7(2e²/h). We use this experimental g-factor to calculate the Kondo temperature. The $\approx 1.5$ times increase in the g-factor, as compared to the bulk, leads to $g_{QPC} \approx 0.66$. Then, $2g_{QPC}\mu_B H|_{H=3\,T} \approx 0.22$ mV. The Zeeman splitting in this field for bulk $|g| = 0.44$ is 0.15mV. According to the formula (1) the difference between these values is $2k_B(T + T_K)$. Now it is easy to calculate $T_K$ when we know the temperature of the experiment.

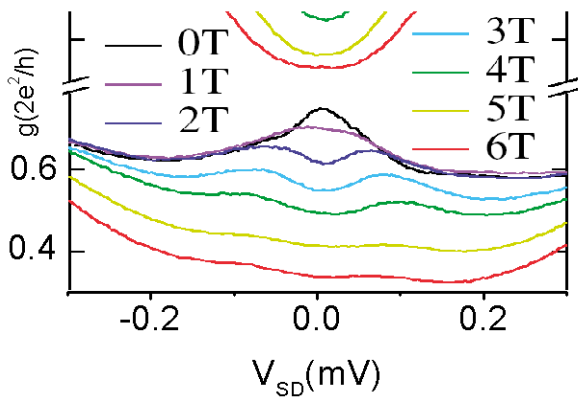

Fig. 1. A fragment of Fig. 2(d) from Ref. [3] showing the evolution of the Kondo feature in an external field parallel to the sample surface.

The dependences $dI/dV(2e^2/h)$ in Fig. 2 (d) (Ref. [3]) permit a similar calculation of the Kondo temperature. The dependences have distinct peaks of the Zeeman splitting (see Fig. 1). For example, we consider the peaks of the Zeeman splitting in the field 2T and 3T, they are most pronounced, and the error in the calculation of $V^{P-P}$ will not be large. The error is nevertheless present if for no other reason than the use of a copy rather than original data. The experimental values in the fields 2 T and 3 T are $V^{P-P}(2T) \approx 0.15$ mV and $V^{P-P}(3T) \approx 0.19$ mV. For these fields the Zeeman splitting $2g\mu_B H$ with $g = 0.44$ is 0.1 mV и 0.15 mV, respectively. Thus, in these fields the experimental $V^{P-P}$ values exceed the Zeeman splitting with $g = 0.44$ on the average by 0.045 mV. According to the formula (1) proposed, this is $2k_B(T + T_K) = 0.045$ mV. Then $T + T_K = 0.265$ K. This value is within the above error.

The Kondo temperature calculated by our formula is in the interval $T_K \approx 0.1 - 0.4$ K found for these compounds from a full width at half maximum (FWHM). The $g$ –factor corresponds to the bulk.

Let us examine another example of calculating the Kondo temperature in QD using the dependence $\Delta V_{DS}(H)$ of Fig. 39 b from Ref. [4], where $\Delta V_{DS}(H) \equiv V^{P-P}(H)$. In these work the $g$-factor is calculated from experimental $V^{P-P}(H)$. The liner approximation in [4] is used for all

$V^{P-P}(H)$ values, including those taken in large magnetic fields. The author of Ref. [4] thus disregards the fact that $V^{P-P}(H)$ measure in too large fields contains an error because the splitting of the peak evolves into a step (fig. 39 a, Ref. [4]), since in this fields $T_{max}$ approaches or exceeds $T_{min}$.

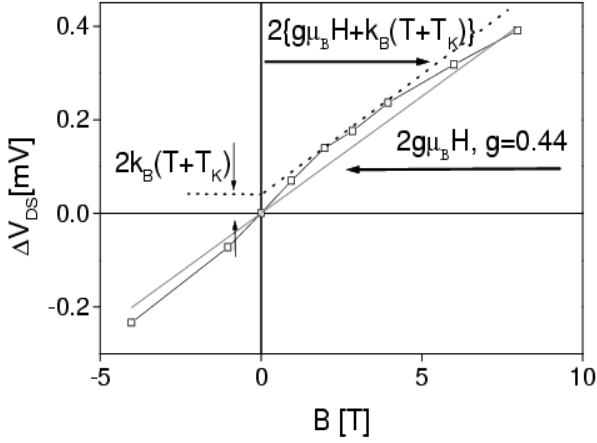

Fig. 2. A copy of Fig. 39 b from Ref. [4]. $\Delta V_{DS} \equiv V_{exp}^{P-P}$ is the distance between two peaks in the magnetic field.

In Fig. 2 the copy of Fig. 39 b from Ref. [4] is supplement with the dependences $2g\mu_B H$ and $2\{g\mu_B H + k_B(T+T_K)\}$ for $g=0.44$. The linear approximation [4] with the slope 0.053 mV/T is omitted. The dependence $2\{g\mu_B H + k_B(T+T_K)\}$ is drawn so that it could pass through the experimental points at which the error in $V^{P-P}$ is smaller. The points taken in large magnetic fields are not included because they transform into steps. Thus, $T+T_K \approx 0.2-0.3$ K, which agrees well with $T_K \approx 0.1-0.4$ K calculated from (FWHM). If we take into account the temperature of the experiment $T=0.09$ K, $T_K$ will be lower.

The example shows that the empirical formula for the Zeeman splitting of the Kondo peak permits an independent calculation of $T_K$ which takes into account the temperature of the experiment and does not change the g-factor for a bulk.

Now we illustrate the validity of the formula for SAT [5]. In this case scattering involves the paramagnetic element Co which forms a single-atom bridge. The two molecules ([Co(**tpy-(CH**2)5-SH)2]$^{2+}$ and [**Co**(**tpy-S**H) 2]$^{2+}$) differ by a 5-carbon alkyl chain within the linker molecules. These molecules were selected because it is known from electrochemical studies that the charge state of the Co ion can be changed from 2+ to 3+ at low energy. Like in QD, the Kondo peak occurs in the differential conductance of the contact. In the magnetic field the peak is split, and the experimental g-factor exceeds $g=2$ predicted for Co.

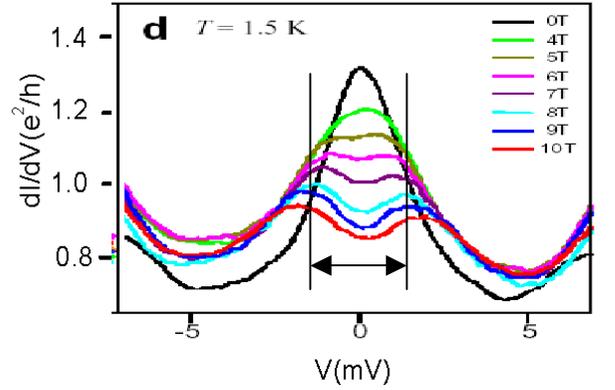

Fig. 3. A copy of Fig. 4 from Ref. [5] supplemented with vertical bars at $V^{P-P} \approx 2.65$ meV in $H$ =8 T.

The above procedure of estimating the Kondo temperature is hardly usable with g-factor of Ref. [5]. In Fig. 3 b of Ref. [5] the dependence $V^{P-P}(H)$ contains experimental points taken in the fields up to $6T$ and gives $g=2.1\pm 0.2$. Two distinct peaks occur only above 6 T (see Fig. 4 d of Ref. [5]). The proposed formula is most efficient when the splitting is well developed. We therefore confine our consideration to $V^{P-P}$ in $H$ =8 T (Fig. 3). As is found from Fig. 4 d in Ref. [5], $V^{P-P} \approx 2.65$ mV. There is an error (note above) in this value but we assume it to be negligible. $T_K$ can be found using our formula and $V^{P-P} \approx 2.65$ mV of the split of Kondo peak at $T=1.5$ K. $2g\mu_B H = 1.84$ mV for

$H = 8$ T and $g = 2$. We then have $2k_B(T + T_K) = 0.81$ meV and $T_K = 3.3$ K. It is much lower than $T_K \approx 20 - 50$ K estimated in Ref. [5].

**Conclusions**

A simple empirical formula is proposed to calculate the Kondo temperature from $V^{P-P}$ of the Zeeman splitting of the Kondo peak in the dependence *dI/dV(V)* for QPC, QD and SAT. In this case the magnetic field should be neither too small, to ensure distinct splitting, nor too large, to keep the split peaks from turning into steps. We do not claim the formula proposed provides a complete description of the Zeeman splitting of the Kondo peak in PCS [1] and QD, GPC, SAT [3,4,5,8-12] Our examples however show that the formula affords a proper consideration of the temperature of the experiment and the Kondo temperature of the Zeeman splitting, which are not negligible. The formula offers a simple technique of estimating the $T_K$ without any fitting parameters, such as e.g., in [18]

I am grateful to I.K. Yanson for useful discussions.